\documentclass[12pt,preprint]{aastex}
\begin{document}

 \title{The Dwarf Novae During Quiescence}

                 \author{Joel A. Urban
                                  and
                             Edward M. Sion}

                  \affil{Dept of Astronomy \& Astrophysics,
                           Villanova University,
                           Villanova, PA 19085,
                     e-mail: jurban@ast.vill.edu,
                               emsion@ast.vill.edu}

\begin{abstract}
We present a synthetic spectral analysis of nearly the entire far 
ultraviolet {\it IUE} archive of spectra of dwarf novae in or near their 
quiescence.  We have examined all of the systems 
for which the signal to noise ratio 
permitted an analysis. The study includes 53 systems of all dwarf nova 
subtypes both above and below the period gap.  The spectra were uniformly 
analyzed using synthetic spectral codes for optically thick accretion 
disks and stellar photospheres along with the best-available distance 
measurements or estimates.  We present newly determined approximate white 
dwarf temperatures or upper limits and estimated accretion rates. 
The implications of 
our study for disk accretion physics and CV evolution are discussed. 
The average temperature of white dwarfs in dwarf novae below the period 
gap is $\sim18,000$K. For white dwarfs in dwarf novae above the period 
gap, the average white dwarf temperature is $\sim$26,000K.  There is a 
flux component, in addition to a white dwarf photosphere, which 
contributes $> 60$\% of the flux in the FUV in 53\% of the quiescent dwarf 
novae in this study. We find that for 41\% of the dwarf novae in our 
sample, a white dwarf photosphere provides $>60$\% of the FUV flux. 
Accretion rates estimated from the FUV alone for the sample of DNe during 
quiescence ranged from $10^{-12} M_{\odot}/$yr to $10^{-10} M_{\odot}/$yr. 
The additional flux component is almost certainly not an optically thick 
accretion disk since, according to the disk instability model, the disk 
should be optically thin and too cool during dwarf nova quiescence to be a 
significant FUV continuum emitter. Among the candidates for the second 
component of FUV light are the quiescent inner disk, a hot accretion belt
at low white dwarf latitudes centered on the equator, 
and hot rotating ring where the 
outer part of the boundary layer (the UV boundary layer) meets the inner disk 
and possibly heats it. The implications of our findings are discussed.
\end{abstract}

\keywords{Stars: white dwarfs, stars: dwarf novae, accretion disks}

\section{Introduction}

Dwarf novae (DNe) are a subclass of cataclysmic variables (CVs), comprised
of a low-mass, main sequence secondary star and a white dwarf (WD)  
primary.  The secondary fills its Roche lobe and serves as the
mass-transferring donor star systems to the accreting white dwarf. DNe are
characterized by their quasi-periodic outburst episodes that are typically
2-6 mag in amplitude.  These outbursts are generally explained as a result
of the mass-transfer from the secondary to the primary.  The secondary
fills its critical Roche lobe, and loses matter through the inner
Lagrangian point.  Since this gas carries substantial angular momentum, it
forms into an accretion ring around the WD. A thermal-viscous instability
known as the disk instability model (DIM) causes the accretion disk to
transition from a cool, quiescent, optically thin state to a stable state
where the disk is much hotter, more luminous, optically thick and
approaching a steady state. This ring becomes broadened into a disk due to
the viscous shear of the matter flowing in the ring.  The disk is heated due
to the release of gravitational potential energy as the accreted material
spirals through the optically thick disk toward the WD surface.  The
energy released can heat the disk to temperatures on the order of 100,000
K.  This heating of the disk accounts for the increase in luminosity
during a dwarf nova (DN) outburst, and should dominate the system's
luminosity in outburst. When the outburst concludes, the disk should,
theoretically, be relatively empty of material in its hotter inner regions
and optically thin.  Unlike the outburst episode, the disk luminosity
should be a negligible contributor to the system's luminosity (Smak 1984; Verbunt 1987).
This period of low mass-transfer is known as quiescence.

The WDs in DNe have been shown to have temperatures that range from about 
9,000 K to as high as 50,000 K (Sion 1999).  These temperatures imply that 
the Planckian peaks of the WDs fall in the far ultraviolet (FUV)  
wavelength range. Because the disk should be fairly empty during 
quiescence, it is expected that the WD will dominate the FUV spectrum of 
the system since, according to the standard theory of disk structure, the 
outer disk is generally too cool to contribute substantially to the far 
UV (Cannizzo 1993). Therefore, FUV analysis of a DN in quiescence should, in principle, 
reveal fundamental characteristics of the underlying WD. However, as we 
discuss below, the definitive identification of the dominant flux source 
of the FUV spectra has been fraught with difficulties.

Numerous authors conducted the earliest optical, FUV and X-ray analyses of 
quiescent dwarf novae (e.g., Patterson 1984; Szkody \& Mattei 1984, 
Patterson \& Raymond 1985). Global studies of FUV spectra of quiescent 
dwarf novae were carried out by la Dous (1991) and Deng, Zhang \& Chen 
(1994), in the former case using {\it IUE} spectra of white dwarfs as 
''templates'' and in the latter study using ratios of FUV color indices. 
However, with the exception of a handful of systems (VW Hyi, U Gem, WZ 
Sge) where FUV studies revealed the Lyman $\alpha$ turnover of a white 
dwarf (Mateo \& Szkody 1984; Kiplinger, Szkody \& Sion 1991; Panek \& Holm 
1984), the FUV continua and any absorption lines were regarded as 
originating in an accretion disk. While this assumption appeared 
physically plausible for the higher mass transfer systems above the period 
gap where denser, sufficiently hot portions of the disk may remain 
optically thick during quiescence, it posed a serious problem for the 
systems below the gap for which the disk instability model (DIM) predicted 
an empty or optically thin inner disk with virtually no flux contribution 
in the FUV. Any progress in understanding the source of the quiescent FUV 
continua and absorption lines was impeded by (1) our lack of knowledge of 
what the FUV emission from optically thin disks would look like and; (2) 
the FUV emission from a white dwarf photosphere and from an (optically 
thick) accretion disk is very difficult to disentangle unless the distance 
to the system is known reliably.

The first attempts to resolve the FUV flux contribution in quiescence with 
actual synthetic spectral models coincided in time with the appearance of 
the Wade and Hubeny (1998) accretion disk model grid and the public 
availability of the model atmosphere codes TLUSTY (Hubeny 1988) and the 
accretion disk code TLUSDISK (Hubeny 1995) which were specifically 
developed for application to hot accretion disks and hot degenerate stars 
with solar abundances. In a series of papers (Lake \& Sion 2001; Nadalin 
\& Sion 2002; Henry \& Sion 2002; Stump \& Sion 2002; Lyons et al. 2002; 
Urban et al. 2003; Sion \& Urban 2003; Sepinsky et al. 2002; Winter \& 
Sion 2003), Sion and co-workers experimented with combinations of high 
gravity photospheres and optically thick accretion disks in synthetic 
spectral fits to the FUV spectra. The choice of optically thick disk 
models was dictated primarily by a lack of anything better or more 
realistic. That being said the Wade and Hubeny (1998) disk models are 
state-of-the-art at this point in time. In most of these cited studies, 
firm conclusions were hampered by the lack of reliable distances and other 
fitting constraints. Nevertheless, it became apparent that for some of the 
systems with reasonably well-determined distances (e.g., SS Aur, RU Peg, 
RX And), a white dwarf photosphere, not an accretion disk, was the dominant 
source of FUV emission (e.g., Lake \& Sion 2001; Sion \& Urban 2003; Sepinsky et 
al. 2002).

This paper represents the most extensive set of DNe in quiescence that has 
been analyzed to date with synthetic spectral fitting.  All sub-types of 
DNe are represented in the sample by at least four members of that 
sub-type.  All of the systems have been cross-referenced with the 
applicable photometric data (i.e., light curves and magnitudes) to verify 
their activity state.  Additionally, the analyses of these systems was 
conducted with a uniform method, utilizing spectral data taken solely from 
the International Ultraviolet Explorer {\it IUE} archives.  Thus, at the very 
least, our approach provides a comparative and statistical sense of how 
these systems differ with respect to each other.  Finally, for the sake of 
comparison, we included parameters derived for a few well-studied systems 
obtained with Hubble Space Telescope {\it HST} and the Far Ultraviolet Space 
Explorer {\it FUSE} and also SU UMa systems analyzed with {\it HST STIS} (Szkody et 
al. 2002) but too faint for {\it IUE}.  Moreover, by having a few systems which 
overlap, we are able to compare the quality of results we obtained from 
{\it IUE} data to those obtained from {\it HST} and {\it FUSE} using similar or identical 
analysis techniques.

\section{ Archival IUE Spectral Data}

The resolution of {\it IUE} was 5 
\AA\ and the spectra were obtained with the Short Wavelength Prime Camera (hereafter SWP) 
encompassing the wavelength range 1170 \AA\ to 2000 \AA.  
All the spectral data obtained from the {\it IUE NEWSIPS} archive are in a low 
activity state, very near or at quiescence.   
All spectra were taken through the large aperture at low dispersion.  The 
Massa \& Fitzpatrick (2000) flux calibration-correction algorithm was 
applied to all the {\it IUE} data used; please see their paper for a description 
of the corrections it makes to the data.  When more than one spectrum with 
adequate SNR was available, the two best spectra were analyzed.  An 
observing log of the observations is given in Table 1, where we list: (1) 
the system name, (2) image number, (3) date of observation (MM/DD/YYYY), 
(4) time of observation, (5) exposure time in seconds, and (6) activity 
state. An asterisk in the last column indicates the quiescence was 
established without ground-based optical light curve data covering the 
time of the {\it IUE} observation.

The activity state of the spectra was determined by examining the AAVSO 
light curves for each system.  Unfortunately, we were unable to 
cross-reference this data, especially for systems not covered by the AAVSO 
data, with data from VSNET (www.kusastro.kyoto-u.ac.jp) during the preparation of this paper due to 
their archives being off-line for an extended period of time.  For systems lacking 
AAVSO light curve data, their activity state was 
assessed based upon either mean photometric magnitudes taken from the 
Ritter \& Kolb (2003) catalogue or from {\it IUE} Fine Error Sensor (FES) 
measurements at the time of the {\it IUE} observation. The FES counts, when 
available, were converted to optical magnitudes to help ascertain the 
brightness state at the time of the {\it IUE} observation. The FES counts could also be
used as consistency checks on our model fitting. This is because the model-predicted optical 
magnitude from a best-fitting model should always be fainter than the observed 
optical magnitude of the system since other sources of optical light (e.g. disk, hot spot, secondary)
should also be contributing. To summarize, quiescence was established 
by the FUV flux level, the FES optical magnitude of the system at the time 
of the observation, the cataloged apparent magnitude in quiescence, 
absence of P-Cygni profiles, presence of emission lines in the data, and 
comparison with spectral data and flux levels for the systems during other 
activity states.  For these systems, the activity states listed in Table 1 
are followed by an asterisk to indicate that they were established without 
observed optical light curve data.

The amount of interstellar reddening for each system was taken from estimates 
in the literature. The three principal sources of reddening were 
the compilations of LaDous (1991), Verbunt (1987) and Bruch and Engel 
(1999). If there was a range of values found, than the value we assume in 
our analyses follows the range shown in column (5) of Table 2, and the 
ranges and assumed values are separated by an arrow.  If only one value 
was listed in the literature, then we adopted that value, and if there was 
none listed, we assumed an E(B-V) of 0.00, and that value is followed by 
an asterisk. The spectra were de-reddened with the IUERDAF routine UNRED.

\section{Synthetic Spectral Fitting Procedure}

\subsection{Distance Constraints on the Fitting}

An integral part of our synthetic spectral analysis is to derive a 
distance from the model fitting and compare it to estimated or derived 
distances to these systems as a consistency constraint on the goodness of 
fit.  The most direct means is to have trigonometric parallaxes such as 
those measured by Thorstensen (2003) for 14 dwarf novae. But dwarf novae, 
unlike other CV subclasses, also offer the advantage of estimating distances 
from correlations between their absolute magnitudes at the peak of 
outburst and their orbital periods. Distances were derived from either 
the absolute magnitude at outburst (M$_{v(max)}$) versus orbital period 
relation of Warner (1995) or from a more recent relationship by Harrison 
et al. (2004) based upon their recent HST FGS parallaxes of dwarf novae. 
The Warner (1995) relation is

\[M_{v(max)} =5.74 - 0.259P_{\rm orb} ({\rm hr})\]   

and the Harrison et al.(2004) relation is 

\[M_{v(max)} =5.92 - 0.383 P_{\rm orb} ({\rm hr})\]. 

At the outset of the modeling process, we applied both of these 
relationships to each system.  Then, we conducted an exhaustive search of 
the literature for previous distance estimates.  If the literature search 
revealed other estimates, then we adopted some reasonable mean based upon 
the different methods used to obtain each distance estimate and the 
distance computed from the two calibrated M$_{v(max)}$ versus P$_{orb}$ 
relations. If no other distance estimates existed, then we simply adopted 
a reasonable mean of the two M$_{v(max)}-P_{orb}$ relations.  If a 
trigonometric parallax was available, then we adopted it for the distance.  
In the case of two systems (U Gem and YZ Cnc), there was more than one 
parallax.  For U Gem, all were identical.  For YZ Cnc, they were slightly 
different, so we took a mean of the parallax values.  In the case of SU 
UMa, the parallax obtained was not well-determined and the distance was
critically dependent on the error estimate adopted (Thorstensen 2003).
Therefore, we combined it in a mean with other distances estimates and the 
M$_{v}$-P$_{orb}$ relations.

\subsection{Synthetic Spectral Fitting}

Model spectra with solar abundances were created for high gravity stellar 
atmospheres using TLUSTY (Hubeny 1988) and SYNSPEC (Hubeny \& Lanz 1995).  
We adopted model accretion disks from the optically thick disk model grid 
of Wade \& Hubeny (1998). After masking emission lines in the spectra, we 
determined separately for each system, the best-fitting white dwarf-only 
model and the best-fitting disk-only model using IUEFIT, our in-house $\chi^{2}$ 
minimization routine.  Taking the best-fitting white dwarf model and 
combining it with the best-fitting disk model, we varied the accretion 
rate of the best-fitting disk model by a small multiplicative factor in 
the range 0.1 to 10 using a $\chi^{2}$ minimization routine called 
DISKFIT (Winter and Sion 2003). Using this method the best-fitting composite white dwarf plus 
disk model is determined based upon the minimum $\chi^{2}$ value achieved 
and consistency of the scale factor-derived distance with the adopted 
distance for each system.  The scale factor, normalized to a kiloparsec 
and solar radius, can be related to the white dwarf radius through:  
$F_{\lambda(obs)} = 4 \pi (R^{2}/d^{2})H_{\lambda(model)}$, where d is the 
distance to the source.

The details of this composite disk plus WD method of analysis and an 
illustration of the accuracy we can obtain in a formal error analysis with 
confidence contours are given in Winter \& Sion (2003). In that paper we 
presented accretion rates of EM Cygni, CZ Ori and WW Ceti in quiescence, 
in which $1\sigma, 2 \sigma$ and $3\sigma$ confidence contours are given. 
In Table 2, we list the final composite fitting results for all systems, 
by column, as the follows: (1) system name, (2) system type (and sub-type, 
if applicable), (3) the instrument with which the system was observed, (4) P$_{\rm orb}$
in fractions of a day, (5) the adopted value or range of E(B-V), (6) distance (in parsecs) 
that was adopted, (7) WD effective temperature (in K), (8) the WD mass 
(in M$_{\odot}$), (9) the inclination angle i, (10) accretion rate (in M$_{\odot}$/yr), (11) percent flux 
contribution of the WD, (12) percent flux contribution of the disk.  
It should be noted that in this table, we include the data from the {\it FUSE} 
and {\it HST} analyses when IUE spectra did not exist. This data is clearly 
denoted in column (3).

There are several distances listed in column (6) of Table 2 that are
followed by either a $\pi$, or a $\pi$ in parentheses.  If the distance is
followed by a $\pi$, that distance is taken from a parallax (or parallaxes
in the case of U Gem).  If the distance is followed by a $\pi$ in
parentheses, than it is an adopted distance that incorporates a parallax
(see the explanation in section 3.1).  All parallaxes are taken from
either Harrison et al. (2004), or Thorstensen (2003).

Since in many cases, the white dwarf masses and 
inclinations were tabulated in Ritter and Kolb with associated error estimates, we adopted 
these values and kept them
fixed in the fitting. While the vast majority of the SWP quiescent spectra were too noisy 
to justify individual formal quantitative error analyses, we estimate the uncertainty in the
derived accretion rates to be approximately an order of magnitude.
For the white dwarf masses inferred from the fitting, we estimate from the 
best-fitting results with photospheres and disks, the uncertainty is roughly $\pm 0.2 M_{\sun}$.
For the inclinations, our estimated uncertainty is $\pm20$ degrees. 

We have illustrated representative fits to the spectra of several dwarf 
novae in quiescence listed in Table 2. The spectra displayed are those of 
the U Gem-type systems X Leo and BV Pup, The Z Cam-type systems KT Per and 
AH Her, the SU UMa-type systems ER UMa, CU Vel, V436 Cen, and short-period system BZ UMa. 
For each of these systems we have displayed the best-fitting accretion 
disk and high gravity photosphere model or a combination of the two.

In Figure 1 we display the {\it IUE} SWP spectrum (SWP15989) of X Leo together
with the best- fitting white dwarf plus disk combination model. X Leo's
spectrum, despite being noisy reveals a possible detection of C III (1175) in
absorption, possible S III + O I (1300)
absorption, an unidentified emission feature at 1420A which is probably an artifact 
due to a radiation event, and the possible
presence of sharp C IV (1550) emission. No other line features are clearly
real. The dashed line in the figure represents the flux contribution of an
optically thick accretion disk while the dotted line represents the white
dwarf flux contribution. From our best fit, we conclude that the white
dwarf is the dominant source of FUV flux, accounting for 65\% of the flux
while the accretion disk source provides 35\% of the FUV flux. The white
dwarf surface temperature is T$_{eff} = 33,000\pm2000$K. The temperature
of the white dwarf in this U Gem-type system is very close to that of the
U Gem white dwarf itself.

In Figure 2, the {\it IUE} spectrum SWP27113 for BV Pup is shown together with 
the best combination disk plus photosphere fit. The spectrum reveals 
strong C IV emission, and absorption features of O I + Si III (1300), Si 
II (1260,1265), the longward wing of Lyman Alpha with a geocoronal 
emission and possible sharp circumstellar or interstellar absorption, 
although the latter may be an artifact. The model fit indicates that BV 
Pup's FUV flux in the SWP range is dominated by an accretion disk which 
accounts for virtually all (95\%) of the FUV light. The indicated 
accretion rate is $\sim 2\times10^{-10} M_{\sun}$/yr.
 
In Figure 3, {\it IUE} SWP17616 for the Z Cam-type dwarf nova KT Per is
displayed along with the best-fitting combination model. The peak to peak
noise in the spectrum of KT Per makes it difficult to identify any line
feature which could be real. Indeed, the only feature identifiable is the
longward wing of Lyman$\alpha$ absorption which is reversed by geocoronal
emission. Our fit is based essentially on the continuum only. The 
continuum slope fitting reveals that KT Per's FUV flux
in the IUE range appears to be completely dominated by the light of an
accretion disk with an accretion rate of $5\times 10^{-11} M_{\sun}$/yr.

In figure 4, the {\it IUE} spectrum (SWP07314) of the Z Cam-type dwarf nova AH
Her is displayed together with the best-fitting combination fit. The
spectrum itself has very strong C IV (1548, 1150) emission but little else
in the way of line features which are real. Our synthetic spectra reveal
that the FUV flux of AH Her is completely dominated by an accretion disk
which accounts for 96\% of the FUV flux. The indicated accretion rate is
$5\times 10^{-10} M_{\sun}$/yr.

In figure 5, the {\it IUE} spectrum SWP54455 spectrum of SU UMa type system ER 
UMa is shown together with the best-fitting model combination. The only 
evident features are C IV (1548, 1550) and Si IV (1393, 1402) in emission. 
The FUV flux is due overwhelmingly to an accretion disk. However, the 
indicated accretion rate is $7\times 10^{-9} M_{\sun}$/yr. This is very 
high for quiescence which raises doubt about the reality of our fit. However, an anonymous 
referee has pointed out that since ER UMa has a very short superoutburst cycle it never really
goes into a deep quiescence. This would help to understand the discrepant accretion rate we have found. 

In figure 6, the {\it IUE} SWP54476 spectrum of SU UMa-type system V436 Cen is
shown together with the best-fitting model combination. The spectrum
reveals several line features: C IV (1550) emission, He II (1640)
emission, Si II (1260, 1265, 1526-33?)  in absorption and O I + Si III
(1300) in absorption, as well as likely Al III (1854-1862 blended)
emission. The FUV flux is due overwhelmingly to an accretion disk.
The indicated accretion rate is $8 \times 10^{-11} M_{\sun}$/yr.  
The white dwarf which contributes 32\% of the light has an upper limit
T$_{eff} = 24,000$K.

In figure 7, the {\it IUE} spectrum SWP54325 of the SU UMa-type system CU Vel is
shown together with the best-fitting model combination. The only features we detect are C IV (1550) 
emission and O I + Si III (1300) absorption. The FUV flux
contributed by the accretion disk is 59\% while the white dwarf 
contributes 41\%. The indicated accretion rate is $6\times 10^{-12} 
M_{\sun}$/yr. The white dwarf has an upper limit T$_{eff} = 21,000$K.

In figure 8, the {\it IUE} spectrum SWP32778 of the SU UMa-type system BZ UMa is 
shown together with the best-fitting model combination. The absence of C 
IV and the strength of N V represent the N/C anomaly in BZ UMa. This 
anomaly is seen in a small fraction of both magnetic and non-magnetic 
systems (see Gaensicke et al. 2003). The top solid curve is the 
best-fitting combination, the dotted curve is the white dwarf spectrum 
alone and the dashed curve is the accretion disk synthetic spectrum alone. 
The white dwarf model has $T_{\rm eff} = 17,000$K, log$g =8$ and the 
accretion disk corresponds to $\dot{M} = 1\times 10^{-11} M_{\sun}$/yr, $i 
= 60^{o}$, and M$_{wd} = 0.55 M_{\sun}$. In this fit, the accretion disk 
contributes 5\% of the far UV flux and the white dwarf 95\% of the flux.

In order to assess the efficacy of our method of identifying how much FUV 
flux, other than from a single temperature white dwarf photosphere, is 
emitted by a quiescent dwarf nova, we have compared our synthetic spectral 
results with the published results obtained by other investigators on the 
same systems. Unfortunately, relatively few estimates of accretion rates 
during dwarf nova quiescence have been published for individual systems. 
However, there have been a number of white dwarf temperature 
determinations. In Table 3, we compare our {\it IUE} temperatures with 
temperatures of independent determinations in other investigations of the 
same systems.

\clearpage

\begin{deluxetable}{lccccl}
\tabletypesize{\tiny}
\tablewidth{0.0pt}
\tablenum{1}
\tablecaption{\sc Observing Log}
\tablecolumns{6}
\tablehead{\colhead{System}&\colhead{SWP\#}&\colhead{Obs.Date}&\colhead{Obs.
Time}&\colhead{Exposure
Time(s)}&\colhead{Outburst-Quiescence Cycle}}
\startdata
SU UMa & 27685 & 2/9/1986 & 0:31:43 & 7199.819&  quiescence, post-sob \\
*** & 35219 & 1/1/1989 & 8:30:09&  6611.633 &  quiescence \\
CUVel & 54235 & 3/27/1995 & 4:25:57 & 14399.768 & quiescence \\
TY Psc & 18613 & 11/20/1982 & 1:55:42&  8399.537 &  quiescence \\
V436 Cen&  54476 & 4/21/1995 & 3:38:50 & 12599.576 & quiescence *  \\
IR Gem &  10791 & 12/10/1980 & 3:50:10 & 7199.819 & near quiescence  \\
EG Cnc&&&&& \\
EK TrA&&&&& \\
OY Car&&&&& \\
HT Cas & 21080 & 9/17/1983&  21:59:12&  22799.844 &  quiescence *  \\
*** & 29626&  7/11/1986 & 12:48:19 & 37499.438 & quiescence *  \\
VWHyi&&&&& \\
EF Peg&&&&& \\
ZCha&&&&&   \\
TLeo & 33699 & 5/25/1988 & 20:04:54 & 2099.48 & near quiescence  \\
UV Per & 21031&  9/13/1983 & 0:49:02 & 12899.813 & quiescence *  \\
YZ Cnc & 7312 & 12/4/1979 & 21:22:34 & 7799.4 73 & quiescence *  \\
A Y Lyr&&&&& \\
ER Uma & 54455&  4/17/1995 & 22:04:26 & 8698.436 & near quiescence, early rise to ob  \\
RZ LMi & 50721 & 5/8/1994 & 12:44:10&  7198.2 & quiescence, post-sob *  \\
SS UMi & 23426 & 7/8/1984&  0:00:25 & 9419.851 & quiescence * \\ 
V1159 Ori&  56861 & 2/27/1996 & 18:41:40 & 6599.755 & near quiescence \\
WZ Sge&&&&&\\
AL Com&&&&& \\
BC UMa&&&&&  \\
VY Aqr&&&&&    \\
WX Cet&&&&&  \\
HV Vir&&&&&  \\
U: And&&&&&    \\
GW Lib&&&&&    \\
BW Scl&&&&&   \\
FS Aur & 14733 & 8/12/1981 & 10:54:53 & 5399.627 & quiescence \\
BZ UMa & 32778 & 1/24/1988 & 20:13:57 & 9599.665 &  quiescence * \\
*** & 32783 & 1/25/1988 & 19:20:12 & 12599.575 & quiescence *  \\
WX Hyi & 7381 & 12/12/1979 & 22:16:17 & 5399.627 & quiescence *  \\
*** & 15918 & 1/1/1982 & 13:37:43 & 7799.473 & quiescence  \\
SW UMa & 23029 &  5/16/1984 & 10:06:05 &  17099.85 & quiescence *  \\
CC Scl & 34427 & 10/6/1988 & 18:47:02 & 7199.819 & quiescence *  \\
V2051 Oph&  21091 & 9/18/1983 & 22:28:46 & 20999.65 & quiescence *  \\
*** & 27811 & 2/28/1986 & 4:55:02 & 20279.574 & quiescence *  \\
VZ Pyx & 44147 & 3/10/1992&  9:12:22 & 5759.666 & quiescence *  \\
*** & 54448 & 4/16/1995 & 22:37:43 & 6298.179 & near quiescence (Szkody \& Silber)*\\ 
U  Gem & 10536&  11/4/1980 & 2:32:48 & 4499.736 & quiescence\\
***  & 11143 & 1/25/1981 & 0.114594907 & 2639.202 & quiescence\\
CN Ori & 32528 & 12/15/1987 & 11:00:10 & 20219.773 & quiescence \\
*** & 32550 & 12/17/1987 & 10:11:46 & 18899.633 & quiescence  \\
RU Peg & 28355 & 5/22/1986 & 4:39:59 & 7619.659 & quiescence  \\
*** & 28683 & 7/16/1986 & 1:04:25 & 6179.505 & quiescence, post-ob \\ 
SS Aur & 16036 & 1/13/1982 & 18:21:55 & 8399.537 & quiescence \\ 
CH UMa & 56270 & 12/6/1995 & 12:15:49 & 13799.704 & quiescence *  \\
TW Vir & 18843 & 12/21/1982 & 4:23:31 & 7199.819 & quiescence *  \\
 BD Pav & 54483 & 4/22/1995 & 2:26:08 & 15599.486 & quiescence *  \\
SS Cyg & 24533 & ll/24/1984 & 4:43:32 & 1319.601 & quiescence  \\
*** & 39907 & 10/21/1990 & 21:45:25 & 1199.588 & quiescence  \\
UU Aql & 21543 & ll/14/1983 & 20:42:55 & 28799.125 & quiescence  \\
EY Cyg & 33428 & 5/2/1988 & 7:56:57 & 10799.793 & quiescence *  \\
CW Mon & &&&&\\
BV Cen & 26623 & 8/16/1985 & 17:17:17 & 14399.768 & quiescence  \\
BV Pup & 27113 & 11/15/1985 & 3:08:22 & 10799.793 & quiescence * \\ 
IP Peg& & & & &   \\
***& & & & &   \\
EX Dra & 45066 & 7/4/1992 & 21:56:54 & 17399.678 & quiescence *  \\
CZ Ori & 16042 & 1/14/1982 & 17:22:22 & 5399.627 & quiescence * \\ 
TT Crt & 41558 & 5/3/1991 & 9:35:09 & 10799.793 & quiescence *  \\
*** & 41575 & 5/5/1991 & 10:02:40 & 10799.793 & quiescence *  \\
FO Aql & 18636 & 11/23/1982 & 11:22:16 & 1679.64 & near quiescence * \\ 
X Leo & 15989 & 1/8/1982 & 14:28:24 & 4799.563 & near quiescence  \\
UZ Ser & 17700 & 8/15/1982 & 18:20:10 & 10799.793 & near quiescence  \\
WW Cet & 24866 & 1/8/1985 & 16:29:20 & 9899.492 &  near quiescence *  \\
AR And &  &  &  &  &    \\
EI UMa & 19038 & 1/19/1983 & 20:58:12 & 5399.627 &  \\
& & & & &   \\
Z Cam & 18844 & 12/21/1982 & 7:42:30 & 2399.717& quiescence \\ 
*** & 37156 & 9/23/1989 & 0:43:39 & 11999.512 & near quiescence \\ 
HL Com & 46283 & ll/16/1992 & 9:26:19 & 4799.563& quiescence \\
*** & 46289 & ll/17/1992 & 9:15:10 & 5399.627 & quiescence  \\
AB Dra & 17595 & 8/5/1982 & 11:12:31 & 1799.652& quiescence  \\
*** & 17596 & 8/5/1982 & 12:23:39 & 1799.652 & quiescence  \\
AH Her & 7314 & 5/12/1979 & 2:38:09 & 3599.844 & quiescence *  \\
VWVul & 52009 & 9/2/1994 & 16:41:03 & 11699.684 & quiescence *  \\
RX And &  &  &  &  &     \\
EM Cyg & 8088 & 2/29/1980 & 15:32:22 & 4499.736 & quiescence *  \\
KT Per & 17616 & 8/7/1982 & 7:08:10 & 6299.518 & near quiescence  \\
TZ Per & 39811 & 10/11/1990 & 22:08:55 & 2399.717 & near quiescence, early rise to ob  \\
 \hline 
\enddata
\end{deluxetable}

\clearpage

\begin{deluxetable}{lcclccccccccc}
\tablewidth{0pc}
\tabletypesize{\tiny}
\tablenum{2}
\tablecaption{\sc Cataclysmic Variables Results}
\tablehead{\colhead{System}&\colhead{Sub-type}&\colhead{Inst.}&\colhead{P$_{orb}$}&\colhead{$E(B-V)$}
&\colhead{d}&\colhead{T$_{eff}$}&\colhead{M$_{\rm wd}$}&\colhead{$i^{°}$}&\colhead{\.{M}}
&\colhead{\% WD}&\colhead{\% Disk}&\colhead{\% Belt}}

\startdata
SU UMa   &SU         &I      &0.07635    &0.00     &270 $\pi$  &28000  & 0.8  & 41  & $9.6\times 10^{-12}$ & 55   &45    &  \\
CU Vel   &SU         &I      &0.0785     &0.00     &200      &21000  & 1.03 & 18  & $6.4\times10^{-12}$ & 40.5 &59.5  &  \\
TY Psc   &SU         &I      &0.06833    &0.00     &238      &16000  & 0.55 & 18  & $1\times 10^{-11}$   & 61   &39    &  \\
V436 Cen &SU         &I      &0.062501   &0.00-0.05&320      &24000  & 0.8  & 41  & $8\times^{-11}$   & 32   &68    &  \\
IR Gem   &SU         &I      &0.0684     &0.00     &250     &&&&&&&\\
EG Cnc   &SU         &H      &0.05997    &0.00     &420      &12300  & 0.6  &     & $1.6\times 10^{-11}$*&      &      &  \\
EK TrA   &SU         &H,I    &0.06288    &0.00-0.05&200  & 18800   & 0.6/0.8 & & $2\times 10^{-11}$ & & &  \\
OY Car   &SU         &H,I    &0.062917   &0.00     &85      & &18000  & 0.685& $1\times 10^{-11}$   &      &      &  \\
HT Cas   &SU         &I      &0.0736472  &0.00-0.10&160  &18000  & 0.8   & 81 &$1\times 10^{-11}$& 96  &4     &  \\
VW Hyi   &SU         &H      &0.0742710  &0.00     &65  & 22000    &14000   & 0.9  &    & $4\times 10^{-12}$* & 64        &36\\
***      &***        &I      &***        &***      &***&&&&&&&\\
EF Peg   &SU         &H,I    &0.0832 *   &0.00     &380  & 16600  & 0.6 & & $2\times 10^{-11}$ *&&&\\
Z Cha    &SU         &H      &0.074499   &0.00     &97       &15700  & 0.59 & 81.7    &$5\times 10^{-10}$*&&    &  \\
***      &***        &I      &***        &***      &***&&&&&&&\\
T Leo    &SU         &I      &0.05882    &0.00     &101 $\pi$ & 16000  & 0.35    &60   & $7\times 10^{-11}$ &  77 &  23&  \\
UV Per   &SU         &I      &0.06489    &0.00     &340   & 20000  & 0.80    &18   & $3.2\times 10^{-11}$ &83 &  17&  \\
YZ Cnc   &SU         &I      &0.0868     &0.00     &265 $\pi$ & 23000  & 0.80    &41   & 1e-10   &24 &  76&  \\
AY Lyr   &SU         &I      &0.0733     &0.00     &440&N/A&&&&&&\\
ER UMa   &SU-ER      &I      &0.06366    &0.00     &460  & 21000  & 0.55 & 60 & $7\times 10^{-10}$ & 12.5 & 87.5&  \\
RZ LMi   &SU-ER      &I      &0.0585 *   &0.00     &940  & 33000  & 1.00 & 41 & $3\times 10^{-10}$ & 10.5 & 89.5&  \\
SS UMi   &SU-ER      &I      &0.06778    &0.00     &1100 & 21000  & 0.55 & 60 & $1\times 10^{-9}$  &  9   & 91  &  \\
V1159 Ori&SU-ER      &I      &0.0621780  &0.00     &260  & 20000  & 0.80 & 60 & $2\times 10^{-10}$ & 10   & 90  &  \\
WZ Sge   &SU-WZ      &Hq     &0.0566878  &0.00     &43.3 $\pi$  & 14800  & 0.90 & 75 & $2\times 10^{-11}$ * & & & \\
AL Com   &SU-WZ      &Hq     &0.056668   &0.00     &800  & 20000  & 0.6 & & $2\times 10^{-11}$ & 61 & 39      &  \\
BC UMa   &SU-WZ      &Hq     &0.06261    &0.00     & 285  & 15200  & 0.60 & & $2\times 10^{-11}$* & 90 & & 10 \\
VY Aqr   &SU-WZ      &Hq,Io  &0.06309    &0.00     &97 $\pi$ & 13500  & 0.8/0.55 & & $5\times 10^{-12}$ * & 86 & & 14  \\
WX Cet   &SU-WZ      &Hq,Id  &0.05827    &0.00     & 187  & 14500  & 0.8/0.55 & & $6\times 10^{-12}$* & 64 & & 36 \\
HV Vir   &SU-WZ      &Hq     &0.05799    &0.00     & 600 $\pi$  & 13300  & 0.60 & & $1\times 10^{-11}$* & & &  \\
LL And   &SU-WZ      &Hq     &0.0550     &0.00     & 760    & 14300  & 0.60 & & $2\times 10^{-11}$* & & &  \\
GW Lib   &SU-WZ      &Hq     &0.05332    &0.00     & 104 $\pi$  & 13300   & 0.80 & & $4\times 10^{-11}$* & 63 & & 37 \\
BW Scl   &SU-WZ      &Hq     &0.054323   &0.00     & 131   & 14800        & 0.60 & &$ 2\times 10^{-11}$* &    & &    \\
FS Aur   &SU(?)      &I      &0.059479   &0.00     & 790  & 25000  & 0.55 & 41 & $4\times 10^{-10}$ & 20 & 80 &  \\
BZ UMa   &SU(?)      &I      &0.06799    &0.00     & 140  & 17000  & 0.55 & 60-75 & $1\times 10^{-11}$ & 95 & 5 &  \\
WX Hyi   &SU(?)      &I      &0.0748134  &0.00     & 320  & 24000  & 1.03 & 60 & $3\times 10^{-10}$ & 8.00 & 92 & \\
SW UMa   &SU(DQ?)    &Iqo    &0.056815   &0.00     & 140  & 16000  & 0.80 & 41 & $3.2\times 10^{-12}$ & 70.5 & 29.5 &  \\
***      &***        &Hq     &***        &***      & 159  & 13900 & 0.6 & & $1\times 10^{-11}$* & 80 & & 20  \\
CC Scl   &SU(IP?)    &I      &0.0584     & 0.00    & 400  & 30000  & 1.21  & 41  & $9.6\times 10^{-12}$ & 38 & 62 &  \\
V2051 Oph&SU(IP?)    &I      &0.062430   &0.00     & 380            & 22000  & 0.80 & 81 & $8\times 10^{-10}$   & 44   & 56   &  \\
VZ Pyx   &SU(IP?)    &I      &0.07332    &0.00     & 320            & 20000  & 0.80 & 60 & $1.6\times 10^{-12}$ & 14.5 & 85.5 &  \\
U Gem    &UG         &H,I,Fo &0.1769062  &0.00-0.05&96 $\pi$  & 30000&1.03&75&$2\times 10^{-11}$&89&11&\\
CN Ori   &UG         &I      &0.163199   &0.00     & 295  & 30000  & 0.80 & 60 & $2\times 10^{-10}$ & 84 & 16 &  \\
RU Peg   &UG         &Iqo,Fq &0.3746     &0.00     & 282 $\pi$  & 50/49000  & 1.21 & 41 & $6.4\times 10^{-12}$ & 84/17.5 & 16 & 82.5  \\
SS Aur   &UG         &Iqo,Fq &0.1828     &0.1      & 201 $\pi$  & 30/31000  & 1.03 & 41 & $1\times 10^{-11}$   & 74  &  26 &  \\
CH UMa   &UG         &I      &0.3431843  &0.00     & 300  & 17000  & 1.21 & 18 & $1.3\times 10^{-11}$ &  3 & 97 &  \\
TW Vir   &UG         &I      &0.18267    &0.00     & 500  & 18000  & 0.80 & 60 & $8\times 10^{-10}$   & 3  & 97 &  \\
BD Pav   &UG         &I      &0.179301   &0.00     & 500  & 18000  & 0.80 & 75 & $6.4\times 10^{-11}$ & 14 & 86 &  \\
SS Cyg   &UG         &I      &0.275130   &0.00-0.04& 166 $\pi$  & 35000  & 12100 & 41 & $5\times 10^{-11}$ & 8 & 92 &  \\
UU Aql   &UG         &I      &0.163532   &0.00     & 345  & 26000  & 0.80 & 18-41 & $3.2\times 10^{-12}$ & & &  \\
EY Cyg   &UG         &I      &0.459326   &0.00     & 450  & 24000  & 1.21 & 18 & $3\times 10^{-12}$ &  34 &  66 &  \\
CW Mon   &UG         &I      &0.1762     &0.00     & 300  & 23000  & 0.80 & 81 & $3\times 10^{-12}$ &  20 &  80 &  \\
BV Cen   &UG         &I      &0.610108   &0.05-0.36& 500  & 25000  & 1.03 & 60 & $6\times 10^{-10}$ & 2.00 & 98 &  \\
BV Pup   &UG         &I      &0.2647     &0.00-0.05& 630  & 26000  & 1.03 & 18 & $2\times 10^{-10}$ & 5.50 &94.5 &  \\
IP Peg   &UG         &I      &0.158206   &0.00     & 200 &&&&&&&\\
EX Dra   &UG         &I      &0.2099372  &0.15     &450   & 22000  & 0.55 & 81 & $9\times 10^{-9}$ &  8 &  92 &  \\
CZ Ori   &UG         &I      &0.214667   &0.00     &260   & 21000  & 0.55 & 18 & $3.5\times 10^{-10}$ &  1 &  99 &  \\
TT Crt   &UG         &I      &0.2683522  &0.00     &500   & 27000  & 0.80 & 60 & $6.4\times 10^{-11}$ & 79 & 21 &  \\
FO Aql   &UG         &I      &none       &0.00-0.10& 99-177  & 21000  & 0.55 & 18 & $2-6\times 10^{-11}$ & 50-74.5 & 25.5-50 &   \\
X Leo    &UG         &I      &0.1646     &0.00     & 350  & 33000  & 1.03 & 41 & $2\times 10^{-11}$ & 65 & 35 &  \\
UZ Ser   &UG(ZC?)    &I      &0.1730     &0.30-0.35& 280  & 99000  & 10300 & 18 & $3.2^{-12}$ &  99 &  1 &  \\
WW Cet   &UG(ZC?)    &I      &0.1758     &0.00     & 190  & 23000  & 0.80 & 60 & $3.2\times 10^{-11}$ &  61 &  39 &  \\
EI UMa   &UG(?)      &I      &0.26810    &0.00     & 730 & & 30000  & 0.80  & $3.2\times 10^{-10}$ & 13.5 & 86.5 &  \\
Z Cam    &ZC(w/SOB?) &I      &0.2898406  &0.00-0.06& 112 $\pi$  &24000  & 1.21 & 60 & $3.2\times 10^{-11}$ &16 & 84 &  \\
HL Com   &ZC(w/SOB?) &I      &0.216778   &0.00-0.10& 575  & 22000  & 10300 & 41 & $9\times 10^{-9}$ & 1.00 & 99 &  \\
AB Dra   &ZC         &I      &0.15198    &0.1      & 400            & 20000  & 1.03 & 81 & $2.9\times 10^{-9}$ & 2.00 & 98 &  \\
AH Her   &ZC         &I      &0.258116   &0.00-0.03& 660 $\pi$   & 29000  & 0.80 & 41 & $3.2\times 10^{-10}$ &  3 &  97 &  \\
VW Vul   &ZC         &I      &0.16870    &0.15     & 605  & 17000  & 0.35 & 41 & $4\times 10^{-9}$ & 2 & 98 &  \\
RX And   &ZC         &Hqd,Io &0.209893   &0.00-0.06& 200  & 35000  & 0.80 & 60 & $3\times 10^{-11}$ & 75 & 25& \\
EM Cyg   &ZC         &I      &0.290909   &0.00-0.05& 350  & 24000  & 1.03 & 75 & $5\times 10^{-11}$ & 8 & 92 &  \\
KT Per   &ZC         &I      &0.1626578  &0.15-0.54& 245  & 26000  & 1.21 & 41-60 & $5\times 10^{-11}$ & 7 & 93 &  \\
TZ Per   &ZC         &I      &0.262906   &0.25-0.30& 435  & 22000  & 1.03 & 41 & $7\times 10^{-10}$ & 1 & 99 &  \\

\enddata
\end{deluxetable}
In column 3, H=HST, I=IUE, F=FUSE, q=quiscence, o=outburst, d=decline. 



\clearpage
\pagestyle{plaintop}

\begin{deluxetable}{lccl}  
\tablecaption{Comparison of WD Temperatures in Dwarf Novae}
\tablenum{3}
\tablehead{
\colhead{System}&\colhead{WD T$_{eff}$ IUE}&\colhead{WD 
T$_{eff}$(other)}&\colhead{Ref.}
}
\startdata        
SS Aur   & 27,000 &           33,000  &       Sion et al. FUSE  \\                               
RU Peg   & 50,000   &          53,000  &        Sion et al. (2004) FUSE\\
WW Ceti  &  23,000    &         27,000  &        Seward et al.(2005) FUSE \\
CU Vel   &  21,000    &         18,500     &     Gaensicke and Koester (1999) IUE\\
U Gem    &  30,000   &          30,000   &       Sion et al.(1998) HST\\
VW Hyi   &  22,000   &          20,000   &       Sion et al.(2001b) HST\\
SS Cyg   &  35,000      &       35,000    &      Holm \& Polidan (1988), Lesniak \& Sion (2003) IUE\\
WZ Sge   &  15,000   &          14,500    &      Sion et al.(1995) HST\\
UU Aql   &  26,000   &          23,000     &     Sion et al. (2005) FUSE\\
HT Cas   &  18,000   &      13,200-18,500  &     Wood, Horne,Vennes(1992) HST\\
EY Cyg    & 24,000   &          24,000     &     Sion et al.(2004) HST + FUSE\\
RX And  &   35,000   &          34,000     &     Sion et al.(2001a) HST\\
EM Cygni &  24,000    &        $<20,000$   &     Welsh et al.(2005) FUSE +  IUE\\
SW UMa   &  16,000      &       16,000     &     Gaensicke and Koester (1999) IUE\\
WX Hyi   &  24,000     &        25,000    &      Long et al.(2005) HST\\
T Leo    &  16,000      &       16,000    &      Hamilton \& Sion (2004) IUE\\           
\enddata
\end{deluxetable}          
\clearpage 
For the most part, we find reasonable agreement between our {\it IUE}
temperatures and the temperatures from other analyses, mostly using {\it HST}
and {\it FUSE}, that are published in the literature. For example, Welsh et
al.(2005) carried out a white dwarf plus accretion disk synthetic spectral
analysis of a combined {\it FUSE + IUE} spectrum of the dwarf nova EM Cygni in
quiescence. They concluded that the FUV spectra were ''dominated by an
accretion disk with a cool (T$ < 20,000$K) white dwarf contributing a
minor but non-negligible flux.'' This is in agreement with the results of
the {\it IUE} analysis of Winter and Sion (2003) and with the re-analysis we
report in this paper. There is also reasonable agreement between our WD
temperatures from {\it IUE} SWP spectra alone and the WD temperatures from
analyses which were carried out on {\it FUSE} spectra alone and combined {\it FUSE}
plus {\it IUE} and {\it FUSE} plus {\it HST} STIS spectra of dwarf novae in quiescence.
Nevertheless, since our white dwarf temperatures are only approximate due
to the low quality and S/N of the {\it IUE} spectra, they should serve only as a
first approximation to the WD temperature. The temperatures in this work,
for the most part, are not intended to be used individually even if better
temperatures are not available with higher SNR spectra from other
spacecraft.

Our method of analysis has proven useful in another way, that is to
indentify those dwarf novae in which the white dwarf star is the dominant
source of FUV flux. In every previous case, up to the time of this
writing, when our {\it IUE} analyses predicted a dominant white dwarf flux
contributor, that prediction has been successfully confirmed with followup
spectroscopy using {\it HST} or {\it FUSE}. This has been the case for SS Aurigae, WW
Ceti, RU Peg, and RX And. On this basis, it would not be surprising if, for those 
systems in Table 2 for which a white dwarf is the dominant contributor, followup 
FUV spectroscopy along with the same distance we had should confirm this.

\subsection{Notes About Specific Systems}

\begin{itemize}
 
\item{CU Vel-- Due to a lack of a normal outburst magnitude in the
literature, the Warner or Harrison relationships could not be used.  We
adopted 200 pc as the distance based upon an estimate by Gaensicke \&
Koester (1999) that corresponded best to the mass estimate.}

\item{ TY Psc- Our inclination estimate from the fitting agrees with
Nadalin \& Sion 2001, but disagrees with Szkody \& Feinswog 1988 (55
deg).}

\item{HT Cas- Wlodarczyk (1986) estimated that the WD contribution to the
FUV spectrum was 8 - 12\%; this is opposite of what we find.}

\item{UV Per- Due to a lack of a normal outburst magnitude in the
literature, the Warner or Harrison relationships could not be used. From
our fitting process, we found that the distance range from the literature
(100 - 263 pc) yielded fits that poorly represented the observed data.  
The model-derived distance from our best fit was 360 pc; therefore, we
have adopted this value as our distance estimate.}

\item{FS Aur- Tovmassian et al. (2003) argue that this DN displays
properties of both SW Sex and intermediate polar (IP) systems, while no
superoutbursts have yet been detected (Andronov 1991). }

\item{BZ UMa- Ringwald et al. (1994) compares this system to V795 Her,
which is thought to be magnetic but doesn't have high-excitation optical
spectra or circular polarization.  Additionally, Kato (1999) finds a
P$_{qpo}$ that is significantly different from the P$_{orb}$, and the
strong X-ray emission of the WD, suggest that this system is an IP.}

\item{WX Hyi- Perna et al. (2003) report single-peaked optical emission
lines, and point out a significant discrepancy between the accretion 
rates estimated from the FUV and X-ray spectra. They also report a broad O VIII emission 
line in the FUV. All of this suggests this might be an SW Sex system.}

\item{SW UMa- Shafter et al.(1986) argued that SW UMa might be an IP
because the soft X-ray flux is strong, the optical spectrum and short
orbital period do not support a high mass-accretion rate, and the optical
and X-ray modulations imply that magnetic accretion is occuring.  They,
along with Patterson (1984), estimate a low accretion rate, which we
confirm, but differs from the high $\dot{M}$ that Szkody et al. (2000)
predict for the system from their analyses.}

\item{CC Scl- Ishioka et al. (2001) propose that this system is a possible
IP based upon the strong X-ray emission, and the ratio of P$_{orb}$ to
P$_{spin}$ in comparison to that of EX Hya.}

\item{V2051 Oph- Our results are not very convincing for this  centrally
eclipsing system. First, we were able to get nearly identical fits using
both a M$_{wd}$ of 0.35 and 0.80 M$_{\odot}$. Second, our fits were less
satisfactory with a 15,000 K WD at either M$_{wd}$ than with a higher 
temperature.  This was surprising since two papers cited a T$_{eff}$ = 15,000 K
(Steeghs et al 2001, Catalan et al. 1998).  Finally, there is the possibility
of a flared disk and/or an iron curtain (Horne et al. 1994), both serving to
confuse the fitting.  Also,
there has been controversy concerning the classification of this system
for nearly twenty years.  Warner \& O'Donoghue (1987) suggest that V 2051
Oph is actually an IP (or low-field polar, as they call it).  Baptista et
al. (1998) explore this possibility with their HST data, but find no
evidence to support the IP hypothesis, but their results are also
inconclusive.}

\item{VZ Pyx- Frequently, this DN is classified as a novalike or an IP.  
However, Hellier et al. (1990) and Patterson (1994) argue that the
variability in the optical and X-ray light curves that are the source of
its questionable classification may not be real.}

\item{CH UMa- Previously, Friend et al. (1990) determined a mass of
$1.95\pm0.30 M_{\sun}$.  The T$_{eff}$/log $g$ fit by Dulude \& Sion
(2003) implies $1.21 M_{\sun}$, which brings the mass down from a neutron
star mass into the WD range.}

\item{UU Aql- For this system, the literature yields distance estimates of
168 - 313 pc (Sproats et al. 1996 and references therein).  Warner's relationship yields 180 pc; Harrison's
relationship yields 298 pc.  Our best fit yielded a scale-factor distance
of 344 pc, and closer distances yielded poorer fits.}

\item{BV Pup- Bianchini et al. (2001) determined a mass estimate of 1.2
M$_{\sun}$, which reduced the mass estimate from $>1.4 M_{\sun}$ (Szkody
\& Feinswog 1988). Our model fits yield a mass estimate of approximately
one solar mass.}

\item{IP Peg- Because of the high inclination, the spectrum is dominated
by disk emission lines.  This, combined with the fact that the SNR of the
spectrum is low, made fitting the spectrum extremely difficult.}

\item{EX Dra- Our mass estimate of $0.55 M_{\sun}$ is somewhat lower than
previous mass estimates from the literature, ranging from $0.66 - 0.75
M_{\sun}$ (Shafter \& Holland 2004, Baptista et al. 2000).}

\item{CZ Ori- Winter \& Sion (2003) adopted a distance estimate of 260 pc,
but their scale factor distance from their best fit was 353 pc.}

\item{FO Aql- This system posed several challenges.  First, there is no
published orbital period.  This prevented us from utilizing the Warner and
Harrison relationships.  Second, there are two very different estimates of
the reddening in the literature.  Szkody (1985) determined there was no
reddening, while Vogt (1983) estimated there was strong reddening.  We
first fitted an unreddened spectrum, then assumed a reddening value of
E(B-V) = 0.1.  The two fits are similar, but the unreddened spectrum
appears to be a better fit, with a lower $\chi^{2}$ value.  Despite the
two different reddening values, some consistencies in the results
appeared.  The mass, inclination, and T$_{eff}$ estimates remained the
same for both fits, and the accretion rate estimates are both very low.  
The scale distances for both differ within a factor of two, and both agree
with the only distance estimate in the literature, $>100$ pc (Berriman et
al 1985).}

\item{UZ Ser- In this re-analysis of the work by Lake \& Sion (2001), who
assumed E(B-V) = 0.0, there are two significant differences we find in the
results.  First, their best inclination was 75 deg. Second, the literature
documents reddening in the range E(B-V) = 0.3 - 0.35(la Dous 1991, Warner
1987, Verbunt et al. 1984).  This made an large difference in the T$_{eff}$
fit. Our estimate of T$_{eff}\geq 99,000$ K could imply the hottest WD in
a DNe yet discovered, a very hot accretion belt with a WD photosphere
nearer to Lake \& Sion's result, or a problem with the quoted reddening
values.  Additionally, Dyck (1988, 1989) suggests that UZ Ser might be a Z
Cam system. This system definitely requires further study.}

\item{WW Cet- This system was classified a Z Cam system by Smak 2002; Ak,
Ozkan, \& Mattei 2002; van Teeseling et al. 1996).  More recently, however,
authors seem to have dropped this classification. This object appears to
be transitional between a nova-like variable and a dwarf nova (Seward et
al. 2005 and references therein).}

\item{EI UMa- Thorstensen (1986) suggests that this system could be a
nova-like or IP due to its strong He II emission and strong hard X-ray
flux.}

\item{HL CMa- Our distance estimate is much greater than any in the
literature.  Neither our scale-factor-derived fits, nor the Warner and
Harrison relationships, support a distance of less than 575 pc.}

\item{AB Dra- Our distance estimate is twice the distance estimated
(Patterson \& Raymond 1985; Berriman 1987), and our inclination is higher
than all values in the literature except one (la Dous 1991).}

\item{SS Cyg- The reddening for SS Cyg is uncertain. Our {\it IUE} temperature
assumes E(B-V) = 0.0. SS Cygni is clearly disk-dominated during quiescence
as first determined with composite WD plus disk models by Lesniak and Sion
(2003). Obviously, with reddening included, the continuum slope fitting
would yield a higher WD temperature.}
\end{itemize}

\section{Discussion}

\indent We have analyzed an archival sample of 53 DNe during their 
quiescence with a synthetic spectral analysis of their {\it IUE} archival 
spectra. This is the largest sample of archival FUV spectra ever analyzed 
with multi-component synthetic spectra for any subclass of cataclysmic 
variables. Our fitting results were further evaluated for consistency 
using flux constraint arguments based upon the theoretical disk 
luminosity, bolometric luminosity of the WD and the observed luminosity, 
using the best available distance estimates [usually the M$_v(max)$ versus 
P$_{orb}$ relations by Warner (1995) and Harrison et al. (2003) based upon 
Hubble FGS parallaxes, and ground-based parallaxes by Thorstensen (2003 
and references therein)].

Two cautions must be emphasized regarding the results in Table 2. First,
as stated earlier, we stress that since our temperatures are only
approximate due to the low quality and S/N of the IUE spectra, they should
serve only as a first approximation to the WD temperature. The
temperatures, for the most part, are not intended to be used individually
even in the absence of better temperatures derived from with higher S/N
spectra with other spacecraft.  Second, the accretion disk models we used
are, at best, only very rough approximations to the quiescent disks in
DNe. The accretion disks during dwarf nova quiescence should not be in a
steady state and should be optically thin except possibly for the cooler,
outer regions of the disks. This being said, applications of steady state
disks to dwarf nova quiescence have appeared elsewhere since our first
composite analyses were published (Sepinsky et al. 2002; Winter and Sion
2003; Sion and Urban 2002; Welsh et al. 2005; Hartley et al. 2005; 
Long et al. 2005). Moreover, despite the well-justified skepticism about our
application of steady state disk models to dwarf nova quiescence,  
''at some level a disk is a disk'' (Wade 2003, private communication). 
For dwarf novae in quiescence, it is not unexpected that some systems may
retain optically thick portions of their disks. The surface density
gradually increases as new matter joins the outer disk where this gas is
added to the reservoir of unaccreted gas remaining from the previous
outburst and must build up to trigger the next outburst. While these disk
models are almost certainly inappropriate for the dwarf novae in
quiescence and nova-like variables in low states, our point is that these
''bad'' disks do represent, by proxy, a different component of system
light, independent of, and in combination with, the accretion-heated
white dwarf models in our analyses. Indeed, it is remarkable that the
synthetic spectra of the accretion disks look suspiciously similar to the
observed FUV spectra seen in several long period dwarf novae at low
inclination.

While one might argue that spectra of lower S/N (typically 2:1 or 3:1 in
these quiescent spectra) like the majority of those in the {\it IUE} archive,
should be ignored if better spectra with other telescopes exist, or might
question the plausibility of the white dwarf plus optically thick disk
combinations we have used, we offer these counter-arguments. First, the
{\it IUE} archive contains the largest sample of FUV spectra of CVs. Second, as
we approach a ''UV-dark'' era with the failure of the {\it HST} STIS, the
uncertain prospect of ever having HST COS operational, and the uncertain
status of {\it FUSE}, it is imperative to fully exploit all possible archival
resources in the FUV. Third, we stress that we have applied a uniform
analysis to the largest, spectroscopic sample of FUV spectra of dwarf
novae obtained with the same telescope, the same instrumental setup, and
the same brightness state, namely dwarf nova quiescence. Fourth, most of
these spectra have never been analyzed or published. Yet, they were
obtained at great expense to US taxpayers and NASA/ESA resources. While
this factor is not a scientific justification and one must always be 
cognizant of their inherent deficiences and limitations, it is incumbent
upon the scientific community to extract new science in any way possible from 
this important archive.

The overall results of our analysis of the IUE archival sample of dwarf
novae in quiescence are stated as follows.

(1)  For the IUE archival sample consisting of 30 dwarf novae above the
period gap and 23 dwarf novae below the gap, we find that the average
white dwarf temperature for the dwarf nova systems in Table 2 below the
period gap is $<T_{eff} > = 18,368$K while for the dwarf novae above the
period gap the average temperature is $<T_{eff} > = 25,793$K. This finding
is consistent with earlier findings for all cataclysmic variable subtypes
collectively, namely that the white dwarfs in cataclysmic variables above
the period gap are hotter and more accretion-heated than those below
the gap (Sion 1991, 1999; Szkody et al.2002; Araujo-Betancor et al.2003).

(2) We believe that we may have uncovered a problem with the current 
understanding of DN quiescence. The disk instability model (DIM) theory 
predicts that the temperature is more or less constant throughout the disk 
and slightly below the critical temperature at which the thermal-viscous 
disk instability is triggered: no FUV radiation should be emitted from the 
quiescent disk (Cannizzo 1993; Hameury 2003 private communication). In contrast, we have 
found that during DN quiescence, the 53 quiescent DNe with usable spectra 
at the lowest flux levels that we analyzed fall into three categories: (a) 
''disk-dominated'' systems where a second FUV flux component, other than a white dwarf,
contributes $>60$\% of the FUV flux. These comprise $\approx 53$\% of the DNe; (b) ''WD-dominated'' 
systems where the WD contributes $> 60$\% of the FUV flux during 
quiescence. These comprise $\approx 41$\% of the DNe and; (c) DNe where 
the WD and ''accretion disk'' contribute roughly equally (between 40\% and 
60\% each). These comprise $\approx$ 6\% of the systems in quiescence. The 
DIM prediction for dwarf novae in quiescence would be satisfied only if 
the white dwarf component is responsible for all of the FUV flux.
We consider it unlikely that with better spectral data and more sophisticated codes
that it will eventually be shown that the white dwarf is responsible for all of the FUV flux 
during dwarf nova quiescence.

(3) For systems above the period gap, 19 out of 30 or 63\% are ''disk-dominated''
and 11/30 or 37\% are WD-dominated. For systems below the gap, 43\% are
''disk-dominated'', 50\% are WD-dominated and 13\% have nearly equal
contributions of WD and disk. If a system is ''disk-dominated'' in the
context of this paper, what does this mean? It means that this ''second
component'' of light could arise from a number of sources among which are
an optically thick portion of the disk, a hot accretion belt at low latitudes centered
on the white dwarf equator, or a hot rotating ring where the outer portion
of the boundary layer (the UV boundary boundary layer) interfaces with the
inner-most accretion disk annulus and possibly heats the inner disk. It is
also important to point out that even an optically thin boundary layer as in 
dwarf nova quiescence can have a non-negligible contribution to the FUV continuum
at the interface between where the outer edge of the boundary layer and the innermost
disk (Popham 1999; Godon and Sion 2005). 

(4) The estimated accretion rates of the {\it IUE} sample of DNe during
quiescence ranged from $10^{-12} M_{\odot}/$yr to $10^{-9.5}
M_{\odot}/$yr. This FUV-derived range is a little broader at the lower end
of the accretion rate range than the range of estimates for dwarf nova
quiescence given by Warner (1995) from his absolute magnitudes and
M$_{v}$-P$_{orb}$ correlations, namely $2\times 10^{-11} M_{\odot}/$yr to
$5\times 10^{-10} M_{\odot}/$yr. This range also agrees with the
accretion rate range derived for 8 dwarf novae in quiescence obtained with
{\it XMM-Newton} X-ray observations by Pandel et al. (2005).  Boundary layer
accretion rates derived by Pandel et al. (2005) from cooling flow models
to the their {\it XMM-Newton} data offer an interesting comparison with the
rough disk accretion rates estimated here. For T Leo, our accretion rate
in Table 2 is virtually identical to the Pandel et al. (2005) boundary
layer value. It is likewise the case for WW Ceti where we estimated
$3\times 10^{-11} M_{\odot}/$yr, the same value as derived by Pandel et
al.(2005). According to the standard accretion theory (Lynden-Bell \& Pringle 1974)
the boundary layer luminosity and accretion disk luminosity should be comparable
for a steady state, optically thick disk but such a disk would more likely
established during an outburst than in quiescence. Pandel et al. (2005)
found that underluminous boundary layers could be ruled out four dwarf novae while
for four other dwarf novae, they found that the boundary layer luminosity
was less than the disk luminosity by factors of 2 to 4.
Our largest deviation from the accretion rates in Pandel et al. occurs for SU UMa
but it was in outburst at the time of the {\it Newton-XMM} observation.

A check of our accretion rates with the few published values for
individual systems also appear not inconsistent. For example, we find,
with {\it IUE} spectra, that SS Cygni's quiescent accretion rate is $\dot{M} =
1\times 10^{-11} M_{\sun}$/yr while Patterson (1984) and Warner (1987)
derived $\dot{M} = 4.75\times 10^{-11} M_{\sun}$/yr from optical and X-ray
measurements. For VW Hyi, the quiescent hard X-ray flux yields an
accretion rate of $\dot{M} = 7\times 10^{-12} M_{\sun}$/yr (van der Woerd
and Heise 1987) whereas using our FUV technique yielded $\dot{M} = 4\times
10^{-12} M_{\sun}$/yr.

Finally, given the sensitivity of the FUV spectral energy distribution and 
line profiles of a disk-accreting CV to the white dwarf mass and the 
inclination angle (through the projected disk area and disk limb 
darkening), it cannot be overemphasized how critical it is to 
independently secure many more accurate WD masses, inclination angles and 
distances to CVs. It is to be hoped that in the years ahead these efforts will 
be greatly intensified.

\acknowledgements

It is a pleasure to express our deepest gratitude to Patrick Godon for
useful comments and to Elizabeth Jewell, Michael Dulude, Ryan Hamilton,
and Scott Engle for their kind assistance in the preparation of this
manuscript. We thank an anonymous referee for numerous useful comments.
This work was supported by NSF grants AST99-01955,
AST05-07514, NASA grants NAG5-12067 and NNG04GE78G and by summer
undergraduate research support from the Delaware Space Grant Consortium.

\clearpage

Figure Captions

\figcaption{The best-fit combination of white dwarf plus accretion disk
synthetic fluxes to the spectrum SWP15989 of the U Gem-type system X Leo
during quiescence. The white dwarf model has $T_{\rm eff} = 33,000$K, log
$g =8.6$ and the accretion disk corresponds to $\dot{M} = 2\times 10^{-11}
M_{\sun}$/yr, $i = 41^{o}$, and M$_{wd} = 1.0 M_{\sun}$. The top solid
curve is the best-fitting combination, the dotted curve is the white dwarf
spectrum alone and the dashed curve is the accretion disk synthetic
spectrum alone. In this fit, the accretion disk contributes 35\% of
the far UV flux and the white dwarf 65\% of the flux.}

\figcaption{The best-fit combination of white dwarf plus accretion disk
synthetic fluxes to the spectrum SWP27113 of the U Gem-type system BV Pup
during quiescence. The white dwarf model has $T_{\rm eff} < 26,000$K, log
$g =8.6$ and the accretion disk corresponds to $\dot{M} = 2\times 10^{-10}
M_{\sun}$/yr, $i = 18^{o}$, and M$_{wd} = 1.0 M_{\sun}$. The top solid
curve is the best-fitting combination, the dotted curve is the white dwarf
spectrum alone and the dashed curve is the accretion disk synthetic
spectrum alone. In this fit, the accretion disk contributes 95\% of
the far UV flux and the white dwarf only 5\% of the flux.}

\figcaption{The best-fit combination of white dwarf plus accretion disk
synthetic fluxes to the spectrum SWP17616 of the Z Cam-type system KT Per
during quiescence. The white dwarf model has $T_{\rm eff} <26,000$K, log
$g = 9.0$ and the accretion disk corresponds to $\dot{M} = 5\times
10^{-11} M_{\sun}$/yr, $i = 41^{o}$, and M$_{wd} = 1.2 M_{\sun}$. The top
solid curve is the best-fitting combination, the dotted curve is the white
dwarf spectrum alone and the dashed curve is the accretion disk synthetic
spectrum alone. In this fit, the accretion disk contributes 93\% of
the far UV flux and the white dwarf 7\% of the flux.}

\figcaption{The best-fit combination of white dwarf plus accretion disk
synthetic fluxes to the spectrum SWP07314 of the Z Cam-type system AH Her
during quiescence. The white dwarf model has $T_{\rm eff} < 29,000$K, log
$g =8.6$ and the accretion disk corresponds to $\dot{M} = 5\times 10^{-10}
M_{\sun}$/yr, $i = 41^{o}$, and M$_{wd} = 1.0 M_{\sun}$. The top solid
curve is the best-fitting combination, the dotted curve is the white dwarf
spectrum alone and the dashed curve is the accretion disk synthetic
spectrum alone. In this fit, the accretion disk contributes 96\% of
the far UV flux and the white dwarf 4\% of the flux.}

\figcaption{The best-fit combination of white dwarf plus accretion disk
synthetic fluxes to the spectrum SWP54455 of the prototype ER UMa during
quiescence. The white dwarf model has $T_{\rm eff} < 21,000$K, log $g
=8.0$ and the accretion disk corresponds to $\dot{M} = 7\times 10^{-9}
M_{\sun}$/yr, $i = 60^{o}$, and M$_{wd} = 0.6 M_{\sun}$. The top solid
curve is the best-fitting combination, the dotted curve is the white dwarf
spectrum alone and the dashed curve is the accretion disk synthetic
spectrum alone. In this fit, the accretion disk contributes 88\% of
the far UV flux and the white dwarf 12\% of the flux.}

\figcaption{The best-fit combination of white dwarf plus accretion disk
synthetic fluxes to the spectrum SWP54476 of the SU UMa-type system V436
Cen during quiescence. The white dwarf model has $T_{\rm eff} = 24,000$K,
log $g =8.3$ and the accretion disk corresponds to $\dot{M} = 8\times
10^{-11} M_{\sun}$/yr, $i = 41^{o}$, and M$_{wd} = 0.8 M_{\sun}$. The top
solid curve is the best-fitting combination, the dotted curve is the white
dwarf spectrum alone and the dashed curve is the accretion disk synthetic
spectrum alone. In this fit, the accretion disk contributes 68\% of
the far UV flux and the white dwarf 32\% of the flux.}

\figcaption{The best-fit combination of white dwarf plus accretion disk
synthetic fluxes to the spectrum SWP54235 of the SU UMa-type system CU Vel
during quiescence. The white dwarf model has $T_{\rm eff} = 21,000$K, log
$g =8.6$ and the accretion disk corresponds to $\dot{M} = 6\times 10^{-12}
M_{\sun}$/yr, $i = 18^{o}$, and M$_{wd} = 1.0 M_{\sun}$. The top solid
curve is the best-fitting combination, the dotted curve is the white dwarf
spectrum alone and the dashed curve is the accretion disk synthetic
spectrum alone. In this fit, the accretion disk contributes 59\% of
the far UV flux and the white dwarf 41\% of the flux.}

\figcaption{The best-fit combination of white dwarf plus accretion disk
synthetic fluxes to the spectrum SWP32778 of the SU UMa-type system BZ UMa
during quiescence. The white dwarf model has $T_{\rm eff} = 17,000$K, log
$g =8$ and the accretion disk corresponds to $\dot{M} = 1\times 10^{-11}
M_{\sun}$/yr, $i = 60^{o}$, and M$_{wd} = 0.55 M_{\sun}$. The top solid
curve is the best-fitting combination, the dotted curve is the white dwarf
spectrum alone and the dashed curve is the accretion disk synthetic
spectrum alone. In this fit, the accretion disk contributes 5\% of the
far UV flux and the white dwarf 95\% of the flux.}
\epsscale{.8}
\clearpage
\plotone{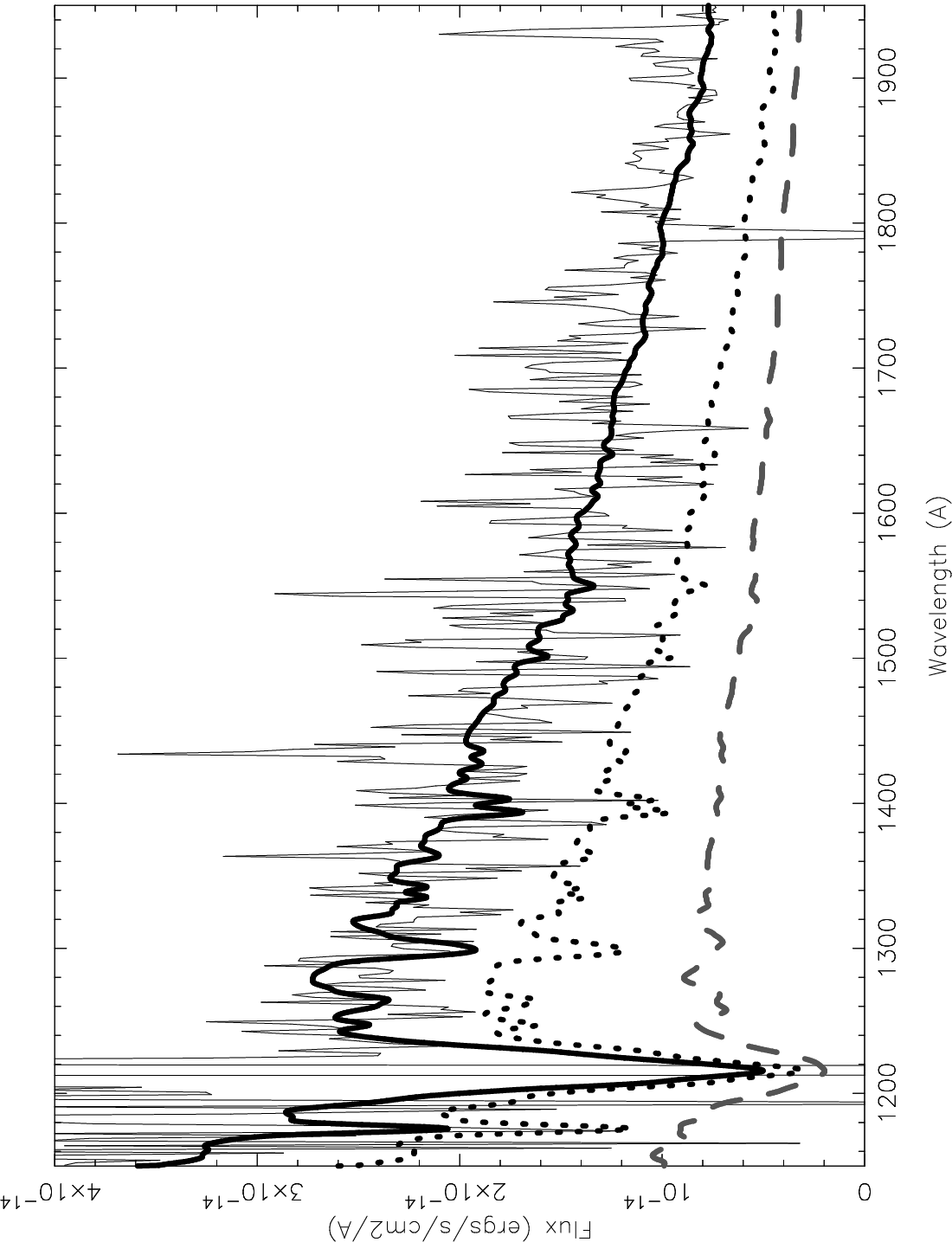}
\clearpage
\plotone{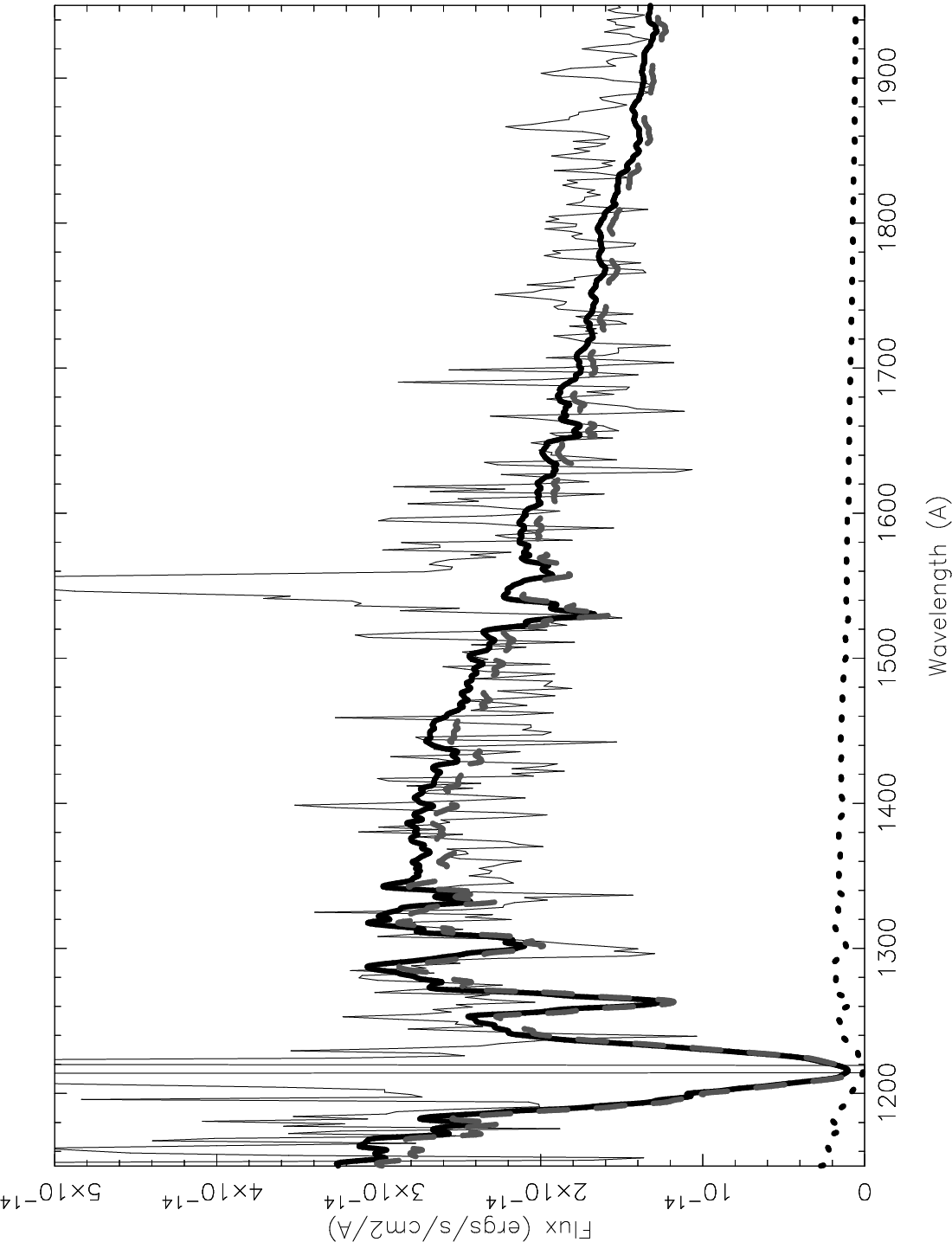}
\clearpage
\plotone{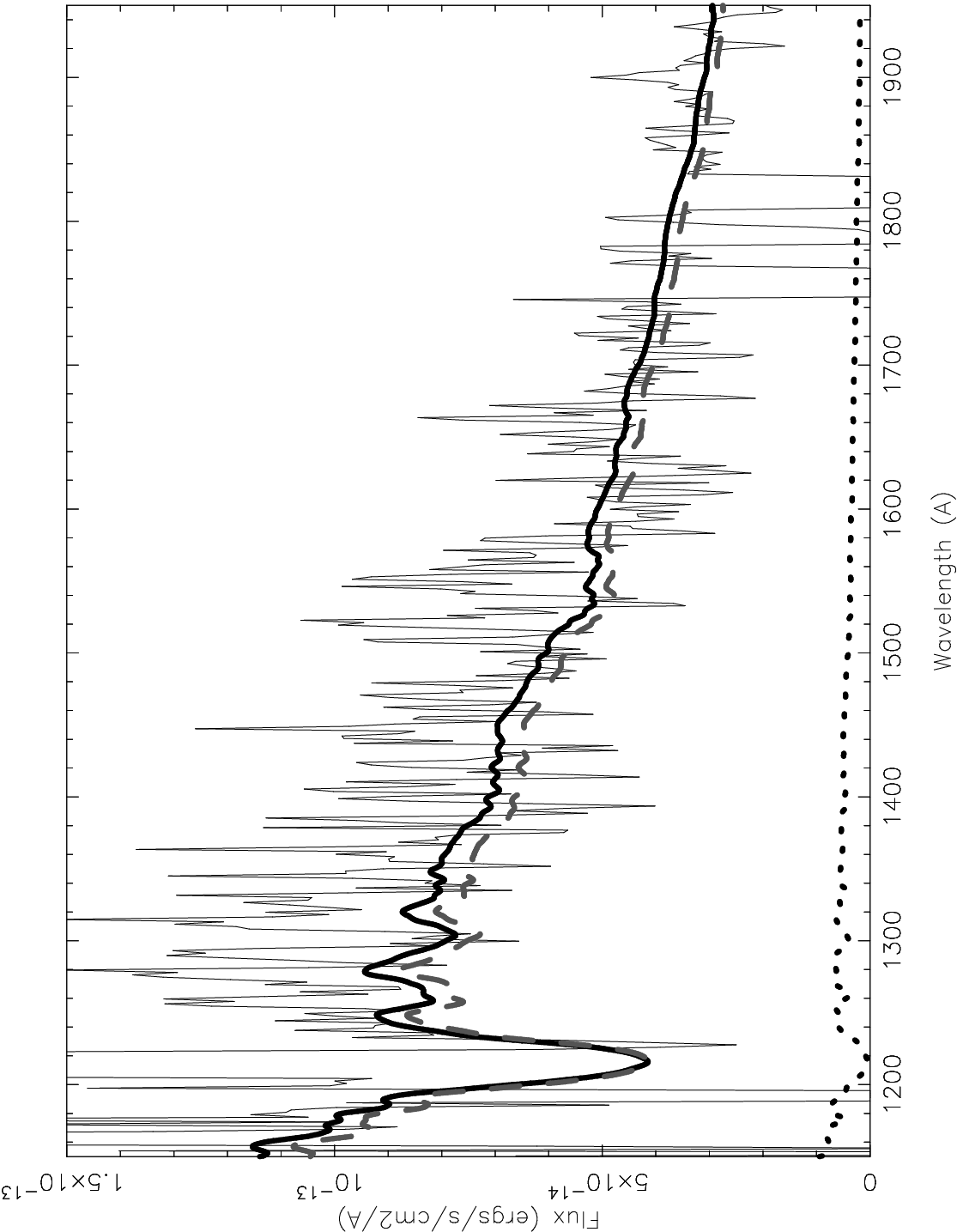}
\clearpage
\plotone{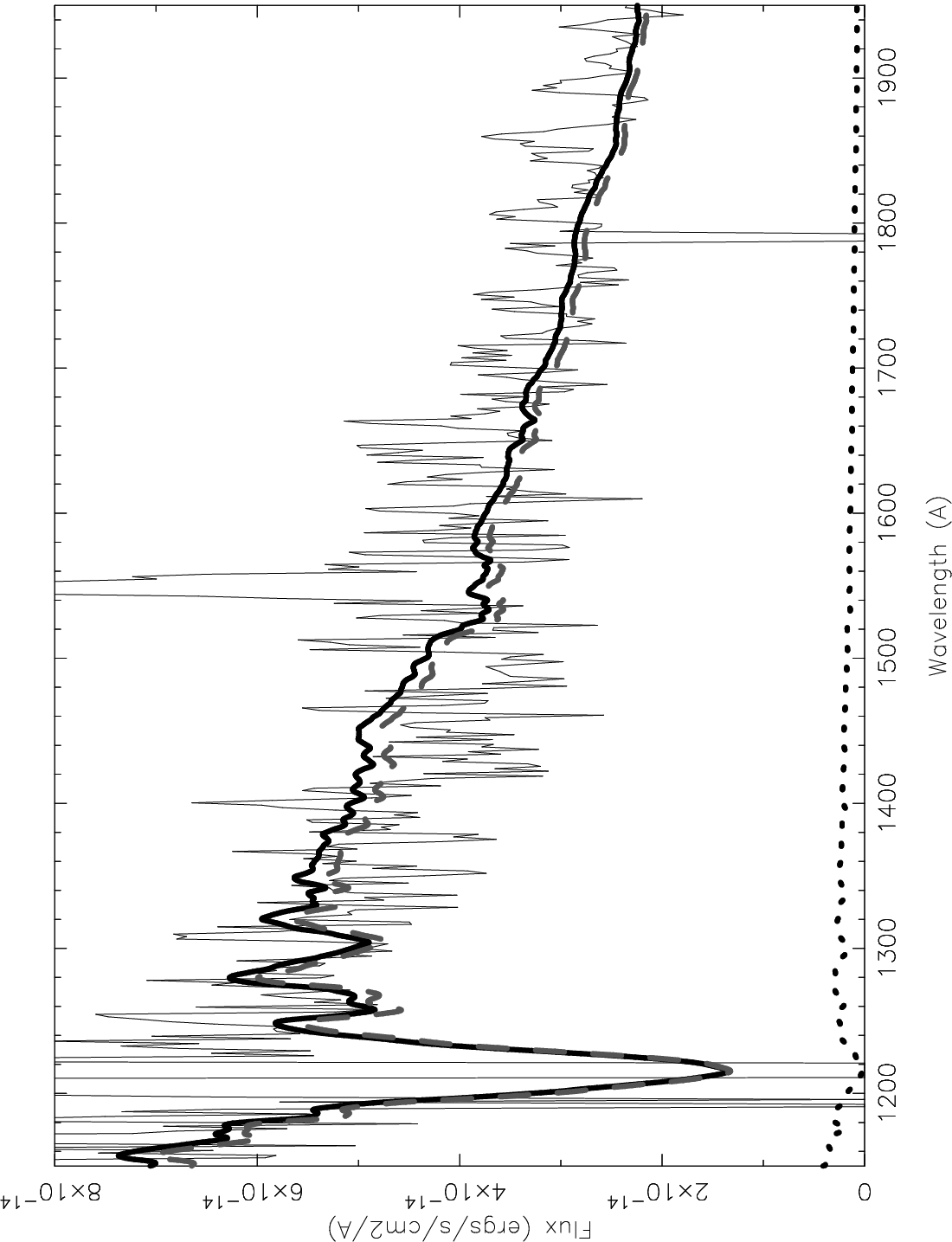}
\clearpage
\plotone{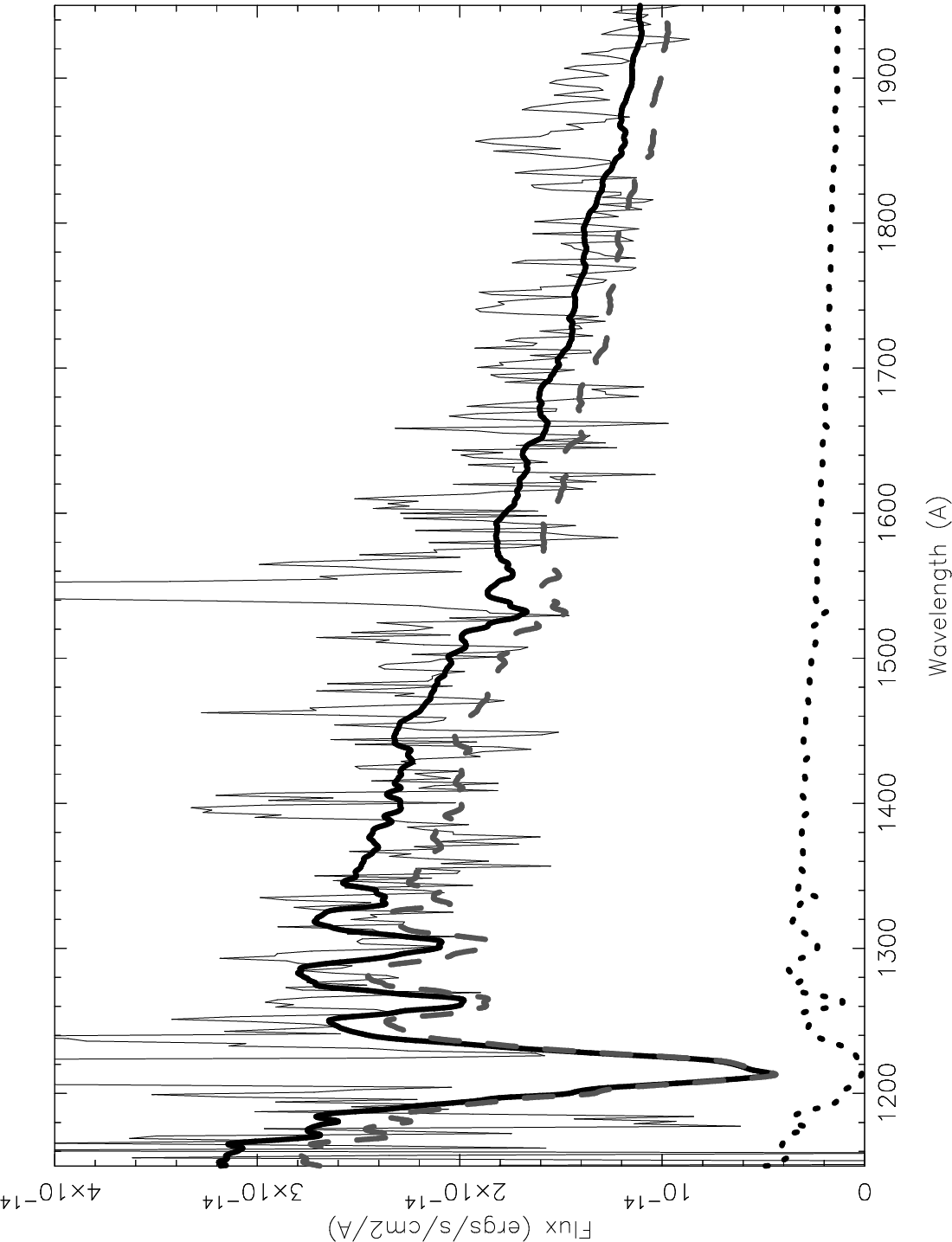}
\clearpage
\plotone{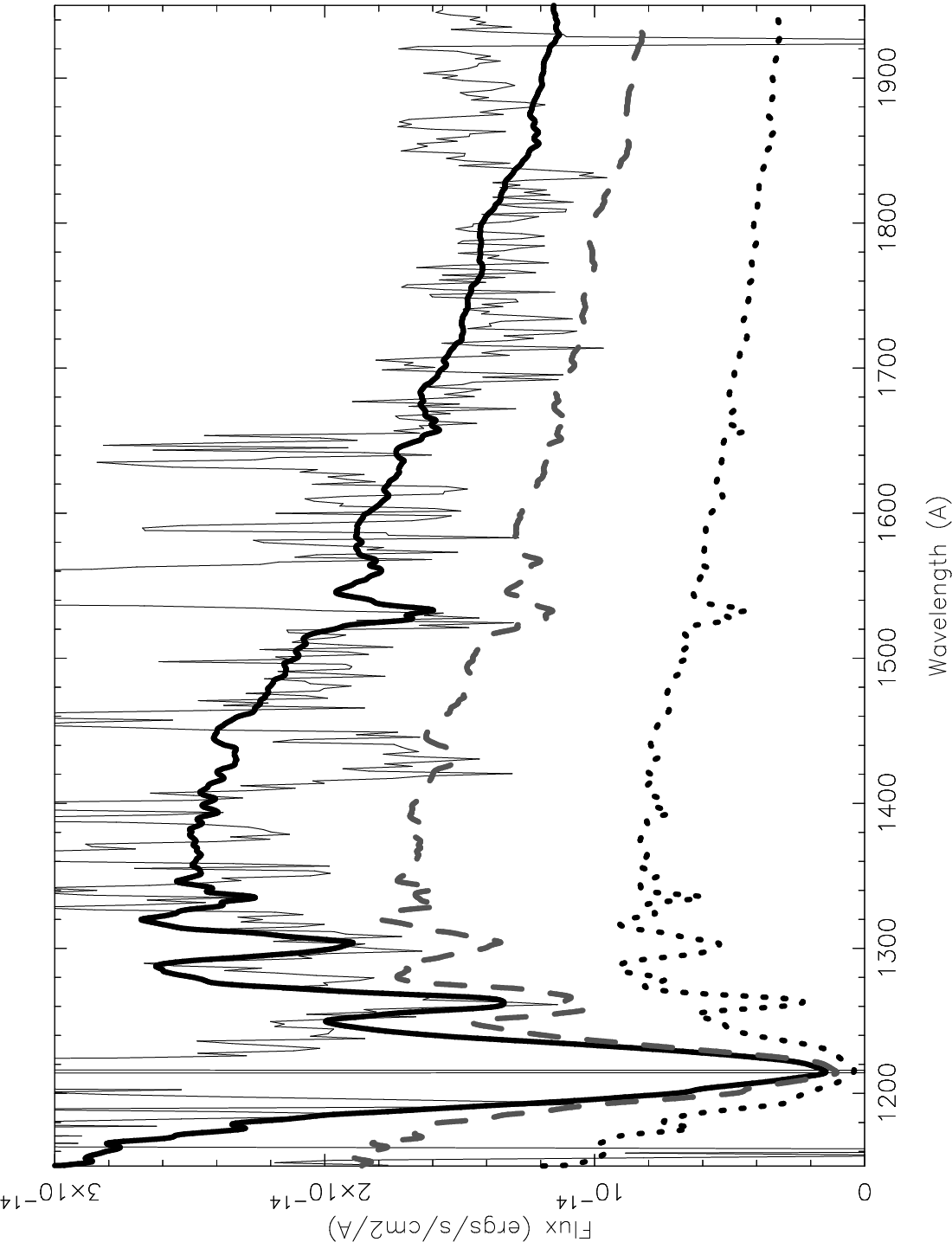}
\clearpage
\plotone{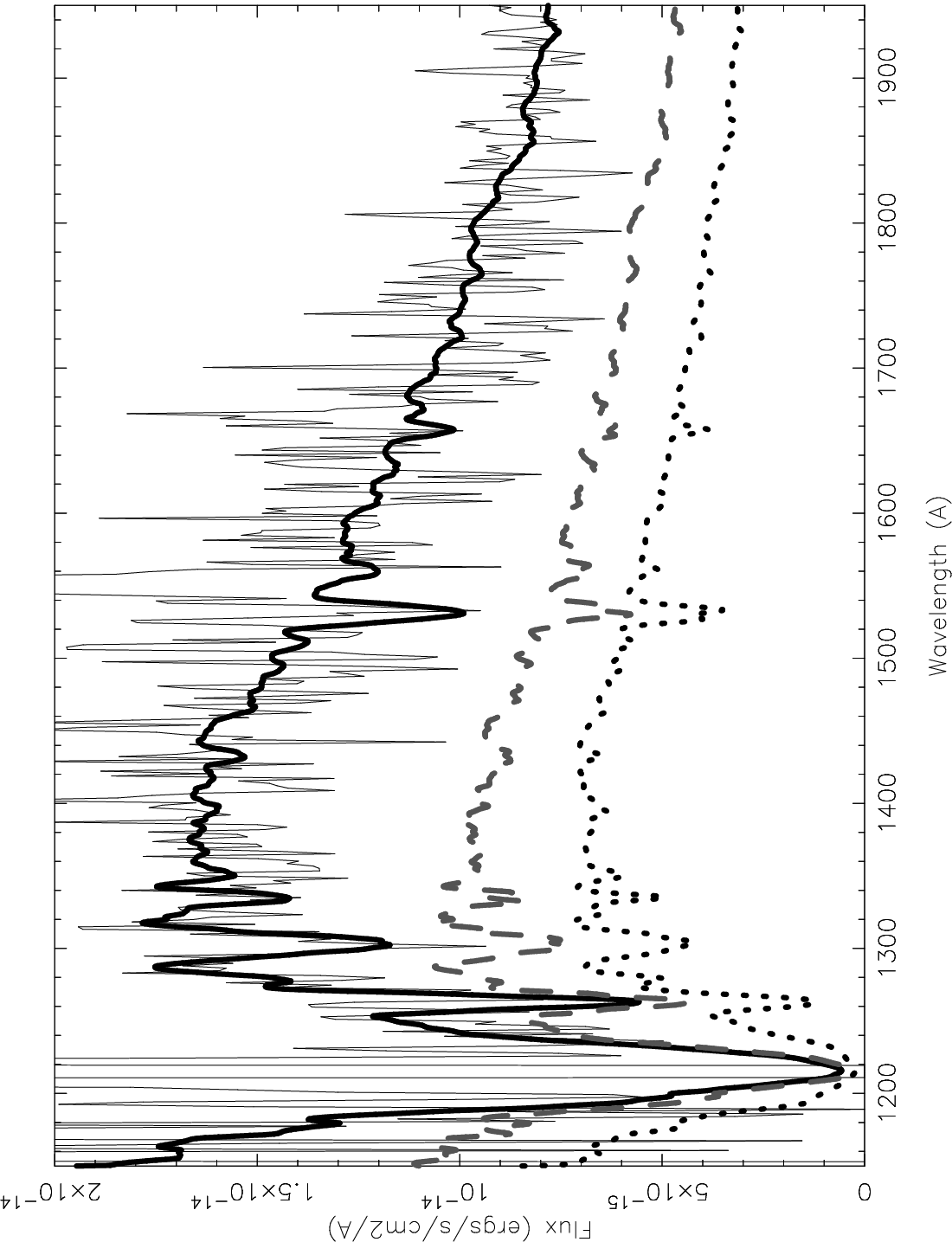}
\clearpage
\plotone{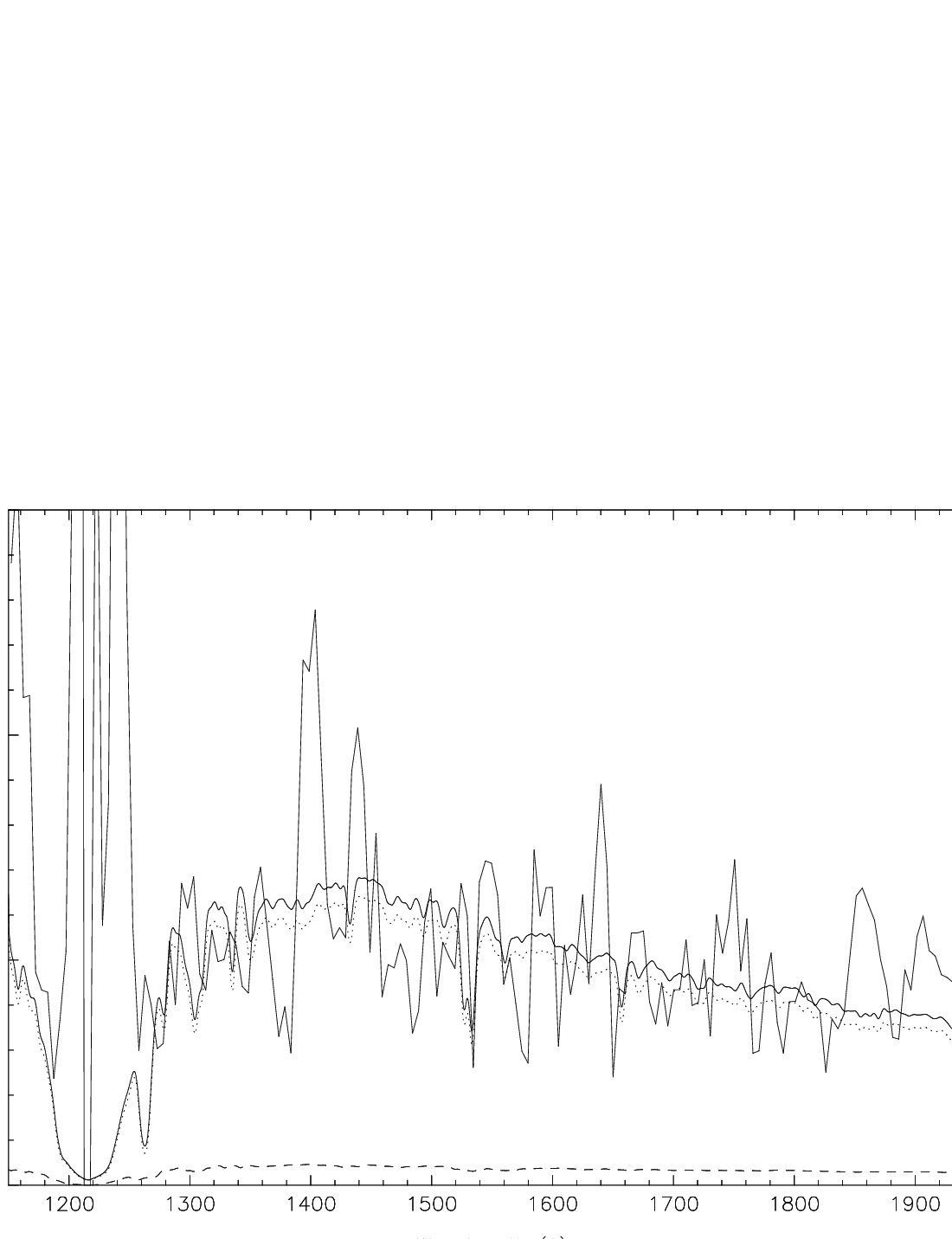}

\end{document}